\documentclass[epj-spec]{svjour}
\def\makeheadbox{}
\usepackage{graphicx}
\usepackage{amsmath}
\usepackage{amssymb} 
\usepackage{bm}

\newcommand{\pdag}{{\phantom{\dagger}}}
\def\bra#1{\ensuremath{\left\langle {#1}\right\vert}}
\def\ket#1{\ensuremath{\left\vert{#1}\right\rangle }}
\newcommand{\braket}[2]{\left\langle{#1}|{#2}\right\rangle}
\newcommand{\up}{\uparrow}
\newcommand{\down}{\downarrow}
\newcommand{\onlinecite}[1]{\cite{#1}}

\newcommand{\TR}{\text{Tr}}
\newcommand{\mydagger}{{\dagger}}
\newcommand{\phdagger}{\phantom{\mydagger}}

\newcommand{\expval}[1]{\langle#1\rangle}

\newcommand{\mydag}{\dagger}

\newcommand{\CC}{\mathcal{C}}

\newcommand{\TC}{\text{T}_{\CC}}

\newcommand{\intC}{\int_\CC}
\newcommand{\tmin}{t_{\text{min}}}
\newcommand{\tmax}{t_{\text{max}}}
\newcommand{\CS}{\mathcal{S}}

\newcommand{\vect}[1]{{\bm #1}}
\newcommand{\Vk}{{\vect{k}}}

\newcommand{\Ks}{{\Vk\sigma}}
\newcommand{\calI}{{\cal I}}
\begin{document}
  \title{New theoretical approaches for correlated systems in nonequilibrium}
  \author{%
    M. Eckstein\inst{1}
    \and
    A. Hackl\inst{2}
    \and
    S. Kehrein\inst{3}
    \and
    M. Kollar\inst{4}
    \and
    M. Moeckel\inst{5}
    \and
    P. Werner\inst{1}
    \and
    F. A. Wolf\inst{4}}
  \institute{%
    Institute of Theoretical Physics, ETH Z\"urich,
    Wolfgang-Pauli-Str. 27, 8093 Z\"urich, Switzerland
    \and
    Institute of Theoretical Physics, Universit\"at zu K\"oln,
    Z\"ulpicher Str. 77, 50937 K\"oln, Germany
    \and
    Arnold Sommerfeld Center for Theoretical Physics and Center for
    NanoScience, Physics Department,
    Ludwig-Maximilians-Universit\"at M\"unchen, Theresienstr. 37,
    80333 M\"unchen, Germany
    \and
    Theoretical Physics III, Center for Electronic Correlations and
    Magnetism, Institute of Physics, Universit\"at Augsburg, 86135
    Augsburg, Germany
    \and
    Max Planck Institute of Quantum Optics, Hans-Kopfermann-Str. 1,
    85748 Garching, Germany
  }
  \abstract{
   We review recent developments in the theory of interacting quantum
   many-particle systems that are not in equilibrium.  We focus mainly 
   on the nonequilibrium generalizations of the flow equation approach 
   and of dynamical  mean-field theory (DMFT). In the nonequilibrium 
   flow equation approach one first diagonalizes the Hamiltonian 
   iteratively,  performs the time evolution in this diagonal basis, 
   and then transforms back to the original basis, thereby avoiding a 
   direct perturbation expansion with errors that grow linearly in time.  In
   nonequilibrium DMFT, on the other hand, the Hubbard model can be
   mapped onto a time-dependent self-consistent single-site problem.
   We discuss results from the flow equation approach for nonlinear 
   transport in the Kondo model, and further applications of this 
   method to the  relaxation behavior in the  ferromagnetic Kondo 
   model and the Hubbard model after an interaction quench. For 
   the interaction quench in the Hubbard model, we have also obtained 
   numerical DMFT results using quantum Monte Carlo simulations. 
   In agreement with the flow equation approach they show that for 
   weak coupling the system relaxes to a ``prethermalized'' intermediate 
   state instead of rapid thermalization. We discuss the description 
   of nonthermal steady states with generalized Gibbs ensembles.}

  \maketitle

  \markboth{}{}

  \section{Introduction}
  \label{intro}

  Strongly correlated electron systems and their rich phase diagrams
  continue to play a central role in modern condensed matter physics.
  Key theoretical developments were the solution of the Kondo model as
  the paradigm for correlated quantum impurity models using the
  numerical renormalization group \cite{Wilson1975}, and the solution
  of the Hubbard model as the paradigm for translation-invariant
  correlated electron systems within dynamical mean-field theory
  (DMFT) \cite{MetznerVollhardt,RMPGeorges}.  Until about ten years ago these
  investigations nearly exclusively focussed on equilibrium or
  near-to-equilibrium (linear response) situations. This was mainly
  due to the fact that experiments probed either equilibrium or linear
  response properties. For example electrical fields applied to bulk
  materials are typically too weak to drive a system beyond the linear
  response regime.

  The experimental situation changed completely in the past ten years
  due to three key developments. First of all, the realization of various
  model Hamiltonians using cold atomic gases paved the way to study
  the real-time evolution of quantum many-body systems and quantum quenches away from equilibrium.
The seminal experiment in this context was the demonstration of collapse 
 and revival oscillations in
     an ultracold gas of rubidium atoms that are described by
 a Bose-Hubbard model quenched from the superfluid to the Mott phase
 by Greiner {\it et al.}  \cite{Greiner}.  
  The second modern experimental development is femtosecond
  spectroscopy \cite{Iwai2003,Perfetti2006,Kawakami2009,Wall2009},
  which permits to study the electron relaxation dynamics in
  pump-probe experiments.  Third is the realization of correlated
  quantum impurities in Coulomb blockade quantum dots
  \cite{GoldhaberGordon,Cronenwett,Schmid}, which can easily be driven
  beyond the linear response regime with moderate applied voltage bias
  \cite{vanderWiel}.

  These experimental developments have led to numerous theoretical
  investigations of nonequilibrium quantum many-body systems in the
  past decade. Still it is fair to say that the 
  nonequilibrium properties of correlated systems are much less understood than their
  equilibrium counterparts. The main reason for this is the lack of
  reliable theoretical methods that can cope with the double challenge
  of strong correlations and nonequilibrium situations. In this paper
  we therefore highlight two such theoretical approaches that have
  been applied successfully to nonequilibrium quantum many-body
  systems, namely the analytical flow equation method \cite{wegner94},
  generalized to nonequilibrium \cite{KehreinBook}, and  nonequilibrium dynamical
  mean-field theory \cite{Schmidt2002a,Freericks2006a}. 
  We discuss how these methods contributed to a first understanding of
  some paradigms of correlated electron physics in nonequilibrium.

  The paper is organized as follows. In Sec.~\ref{sec:2} we introduce
  the flow equation approach for analytic diagonalization of quantum
  many-body systems, especially its generalization to nonequilibrium
  real-time evolution problems. Sec.~\ref{sec:3} deals with the
  application of this approach to the ferromagnetic Kondo model, which
  can thereby be solved in a controlled way even for asymptotically
  large times. The results are in very good agreement with numerical
  methods \cite{hackl09b} and also establish a key mechanism for
  thermalization after an interaction quench in the Hubbard model.
  The time evolution of the Hubbard model is then analyzed in
  Sec.~\ref{sec:floweqHubbard} using the flow equation method.
  Sec.~\ref{sec:Kondotransport} contains an application of the flow
  equation method to a second important class of nonequilibrium
  problems, namely to transport beyond the linear response regime.  In
  Sec.~\ref{sec-dmft} we briefly review nonequilibrium DMFT, which
  maps the lattice system onto an effective single-site problem with a
  self-consistency condition. 
We show how this self-consistency condition can be reduced to a 
 single equation in some cases, which can reduce the numerical 
 effort for the solution of DMFT, in particular in the absence 
 of translational invariance in time.
 In Sec.~\ref{dmftquench} we
  discuss DMFT results for interaction quenches in the Hubbard model,
  which agree very well with the results from the flow equation method
  for small values of the Hubbard interaction
  (Sec.~\ref{sec:floweqHubbard}). For quenches to intermediate
  Hubbard interaction the system thermalizes on short time scales,
  which shows that correlated systems in isolation can reach a new
  equilibrium state, not because of a coupling to external baths but
  due to the interactions between particles.  In
  Sec.~\ref{chainquench} we compare this behavior to that in a
  special solvable Hubbard model, for which the system tends to a
  nonthermal state due to its integrability, i.e., the conservation of
  many constants of motion. Finally in Sec.~\ref{gge} we discuss the
  concept of generalized Gibbs ensembles, which can make statistical
  predictions both for integrable and nearly integrable Hamiltonians.

  \section{Flow equation approach to quantum time evolution}
  \label{sec:2}

  Stationary eigenstates of a many-particle Hamiltonian are
  fundamental for the discussion of quantum many-particle systems in
  equilibrium.  A typical class of {\em nonequilibrium} situations are
  systems that are prepared in a quantum state $| \Psi_i\rangle$ which
  is not an eigenstate of the Hamiltonian $H$ that drives its time
  evolution.  In this case, observables that do not commute with $H$
  will generally become time-dependent.  The canonical way to evaluate
  the real-time evolution of such observables is the Keldysh
  technique. One of the notorious difficulties with this approach is
  that often the limit of weak interactions does not commute with the
  limit of long times.  A partial sum over certain diagrams is usually
  not sufficient to guarantee a controlled approximation in the limit
  of long times.

  Another approach for calculating time-dependent observables is
  Heisenberg's equation of motion for an observable~$A$,
  \begin{equation}
    \frac{dA}{dt}=[H,A] \ .
    \label{heisenberg}
  \end{equation}
  Again, a direct perturbative expansion of this equation of motion,
  e.g., in the interaction strength, is usually not controlled in the
  limit of long times. However, one might think of changing the basis
  representation of Eq. (\ref{heisenberg}).  Ideally, this basis
  representation would transform $H$ into a non-interacting form. This
  idea is precisely at the heart of the {\em flow equation approach},
  invented by Wegner in 1994 \cite{wegner94}, and independently by
  G{\l}azek and Wilson
  \cite{glazek93,glazek94}.

  The main concept is to implement this transformation as a sequence
  of infinitesimal unitary transformations that allows to separate
  energy scales during the process of transformation.  This is
  achieved by introducing an appropriate generator $\eta(B)$ that
  parameterizes a family of unitarily equivalent Hamiltonians.  By
  solving the differential equation
  \begin{equation}
    \frac{dH}{dB}=[\eta(B),H(B)]
    \label{flowH}
  \end{equation}
  with the initial condition $H(B=0)=H$, a unitary equivalent family
  of Hamiltonians is constructed. Wegner's ingenious choice of a {\em
    canonical} generator is to construct it as the commutator of the
  diagonal ($H_0$) with the non-diagonal part ($H_{\rm int}$) of the
  Hamiltonian,
  \begin{equation}
    \eta(B)\stackrel{\rm def}{=} [H_0(B),H_{\rm int}(B)]
    \ .
    \label{defeta}
  \end{equation}
  This ensures that $H(B)$ becomes more and more energy-diagonal with
  an effective band width $\Lambda_{\rm feq}$ during the flow. One can
  easily verify the identification $\Lambda_{\rm feq}\propto
  B^{-1/2}$.  Generically one reaches a non-interacting diagonal form
  in the
  limit $B\rightarrow \infty$ \cite{wegner94}.

  \begin{figure}
    \begin{center}
      \resizebox{0.65\columnwidth}{!}{
        \includegraphics{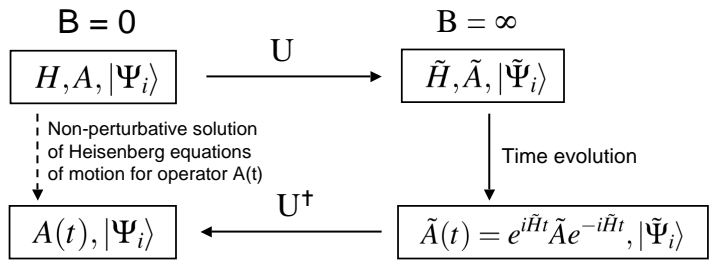}
      }
    \end{center}
    \caption{The forward-backward transformation scheme induces a non-perturbative
      solution to the Heisenberg equation of motion for an operator $A$. $U$ denotes
      the full unitary transformation that relates the $B=0$ to the $B=\infty$ basis \cite{hackl08a}.}
    \label{fig:1}
  \end{figure}

  Intuitively, the basis defined by constructing the non-interacting
  normal form of the Hamiltonian should be much more suitable to solve
  Heisenberg's equation of motion. This has been used in numerous
  examples for evaluating equilibrium correlation functions on all
  energy scales using flow equations \cite{KehreinBook}.  At the same
  time it also allows the calculation of nonequilibrium real-time
  evolution problems.  Indeed, the analogy of constructing normal
  forms of interacting Hamilton functions in order to integrate
  Hamilton's equation of motion is a very successful concept in
  classical mechanics, known as {\sl``canonical perturbation theory''}
  \cite{goldstein02}. Based on the flow equation approach, this idea
  can be directly applied to quantum many-body systems
  \cite{hackl08a,hackl09c}.  The general setup is described by the
  diagram in Fig.~\ref{fig:1}, where $| \Psi_i \rangle$ is some
  initial non-thermal state whose time evolution one is interested in.
  Here and in the following, we describe operators and coupling
  constants in the $B=\infty$ basis by a tilde.  In order to study the
  real-time evolution of a given observable $A$ that one is interested
  in, the observable is transformed into the diagonal basis by solving
  the differential equation
  \begin{equation}
    \frac{dO}{dB}=[\eta(B),O(B)]
    \label{opflow}
  \end{equation}
  with the initial condition $O(B=0)=A$. The key observation is that
  one can now solve the real-time evolution with respect to the
  energy-diagonal $\tilde{H}$ exactly, thereby avoiding any errors
  that grow proportional to time (i.e. secular terms): this yields
  $\tilde{A}(t)$. Now since the initial quantum state is given in the
  $B=0$ basis, one undoes the basis change by integrating
  Eq.~(\ref{opflow}) from $B=\infty$ to $B=0$ (backward
  transformation) with the initial condition
  $O(B=\infty)=\tilde{A}(t)$.  One therefore effectively generates a
  new non-perturbative scheme for solving the Heisenberg equation of
  motion for an operator, $A(t)=e^{iHt}A(0)e^{-iHt}$, in exact analogy
  to canonical perturbation theory.  In a first successful
  application, this approach has been applied to dissipative quantum
  systems \cite{hackl09c,hackl08a}. In this paper, we will discuss
  recent work on the ferromagnetic Kondo model (Sec.~\ref{sec:3}),
  the fermionic Hubbard model
  (Sec.~\ref{sec:floweqHubbard}) and on steady state transport through Kondo dots (Sec.~\ref{sec:Kondotransport}).

  Usually, the implementation of the flow equation approach relies on
  approximations that allow to truncate the hierarchy of flowing
  interaction terms generated during the transformation of an
  interacting many-particle Hamiltonian and all observables. Systems
  that are accessible by perturbative RG are ideally suited for that
  purpose, since they allow for a controlled expansion around a
  weak-coupling fixed point.

  \section{Ferromagnetic Kondo model}
  \label{sec:3}

  A well-known example for a weak-coupling many-particle system can be realized in 
  the ferromagnetic regime of the Kondo model 
  \begin{equation}
    H =\sum_{\bm{k}\sigma} \varepsilon_{\bm{k}} c_{\bm{k}\sigma}^\dagger c_{\bm{k}\sigma}^\pdag 
    + \sum_{\bm{k},\bm{k}^\prime} J_{\bm{k}^\prime \bm{k}}^{\parallel } S^z s_{\bm{k}^\prime \bm{k}}^z 
    + \sum_{\bm{k},\bm{k}^\prime} J_{\bm{k}^\prime \bm{k}}^{\perp} (S^+ s_{\bm{k}^\prime \bm{k}}^- + S^- s_{\bm{k}^\prime
      \bm{k}}^+) \ ,
    \label{hkondo}
  \end{equation}
  where $S^\pm=S^x\pm iS^y$ and likewise for the conduction electron
  spin densities.  In the following, we consider constant {\em
    ferromagnetic} exchange couplings with $|J^\perp|\leq J^{\parallel}$.
  The canonical generator is immediately obtained from
  $\eta(B)=[H_0,H_{\text{int}}(B)]$, where the flowing interaction
  $H_{\text{int}}(B)$ is parametrized by the flowing couplings
  $J_{\bm{k},\bm{k}^\prime}^\parallel(B)$ and
  $J_{\bm{k},\bm{k}^\prime}^\perp(B)$. At the Fermi surface (with
  $J_{\bm{k}_{F},\bm{k}_{F}}^{\perp ,\parallel}\equiv J^{\perp ,\parallel}$ and the
  density of states $\rho$), it can be shown that these couplings
  reproduce the usual poor man's scaling equations \cite{anderson70}
  \begin{eqnarray}
    \frac{dJ^\parallel}{d \ln\Lambda}   &=&- \rho (J^{\perp})^2 \ , \nonumber\\ 
    \frac{dJ^\perp}{d \ln \Lambda}&=&-\rho J^\perp  J^\parallel \ .
    \label{pms}
  \end{eqnarray}
  For ferromagnetic couplings, the non-interacting fixed point of Eq.
  (\ref{pms}) is stable and allows for a controlled perturbative
  expansion of all flow equations used to transform $H$ and $S^z$.
  This example allows to analytically implement the approach outlined
  in Sec.~\ref{sec:2} to calculate the time-dependent impurity
  magnetization $\langle S^z(t)\rangle$ with respect to the initial
  state
  \begin{equation}
    | \Psi_i \rangle= |\uparrow \rangle \otimes | \text{FS} \rangle \ .
  \end{equation}
  The state $|\Psi_i \rangle$ has the physical interpretation of an
  initially decoupled system of a polarized impurity spin $|\uparrow
  \rangle$ and a non-interacting Fermi sea in equilibrium, $|
  \text{FS} \rangle$. The remaining technical steps consist of
  transforming the operator $S^z$ according to Fig.~\ref{fig:1}, in
  order to evaluate the observable $\langle S^z(t)\rangle$.

\begin{figure}
    \begin{center}
      \resizebox{0.65\columnwidth}{!}{
        \includegraphics{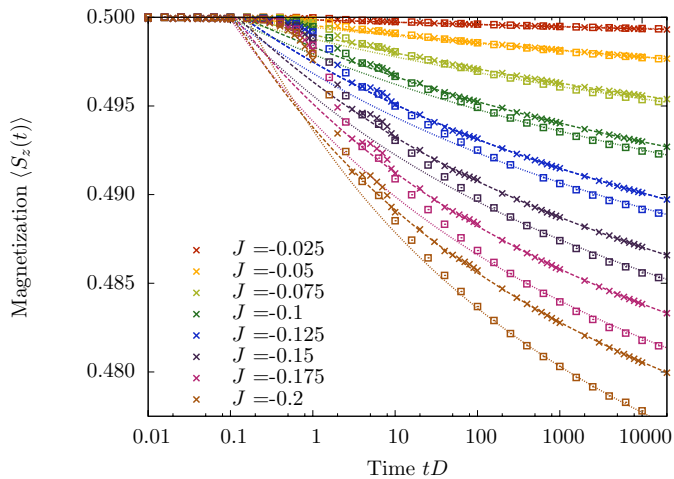}
      }
    \end{center}
    \caption{Time-dependent magnetization in the {\em isotropic}
      ferromagnetic Kondo model. $\Box$ denotes the flow equation
      data points, which agree very well with time-dependent numerical
      renormalization group calculations depicted by $\times$
      \cite{hackl09b}. The deviations for larger values of the
      coupling constant $J$ can be understood from the perturbative
      nature of the flow equation calculation \cite{hackl09a}.  Using
      our analytical result we fitted the flow equation data against
      $\langle S^z(t)\rangle=(1+aJ+[\ln(t)-1/(cJ)]^{-1})/2$ using $a,
      c$ as fit parameters (lines).}
    \label{fig:2}       
  \end{figure}

  The impurity magnetization is obtained as follows.  It is
  straightforward to work out the ansatz for the flowing spin operator
  as
  \begin{equation}
    S^z(B)=h(B)S^z+\sum_{\bm{k}\bm{k}^\prime}\gamma_{\bm{k}^\prime \bm{k}}(B):(\vect{S} \times \vect{s}_{\bm{k}^\prime \bm{k}})^z:\ .
    \label{ansatzb}
  \end{equation}
  In the limit $B \rightarrow \infty$, the coupling constant $h(B)$
  approaches the value $h(B \rightarrow \infty) = 1+ \rho
  J/2+\mathcal{O}(J^2)$ if $J^{\parallel}=J^\perp$. The value $h(B
  \rightarrow \infty)/2$ is equal to the equilibrium impurity
  magnetization in presence of an infinitesimal Zeeman term $0^+S^z$
  and has been calculated to the same accuracy by Abrikosov
  \cite{abrikosov70} by a summation over parquet diagrams. According
  to our strategy for the solution of Heisenberg's equation of motion,
  the nonequilibrium magnetization follows from an ansatz
  \begin{equation}
    S^z(B,t)=h(B,t)S^z+\sum_{\bm{k}\bm{k}^\prime}\gamma_{\bm{k}^\prime \bm{k}}(B,t):(\vect{S} \times \vect{s}_{\bm{k}^\prime\bm{k}})^z:\ .
    \label{ansatzt}
  \end{equation}
  Obviously, the flow equations for this operator have the same form
  as the time-independent flow equations for the ansatz in Eq.
  (\ref{ansatzb}). It is furthermore trivial to determine the initial
  condition as $S^z(B\rightarrow \infty,t)=h(B\rightarrow
  \infty)S^z+\sum_{\bm{k},\bm{k}^\prime}
  \gamma_{\bm{k}^\prime\bm{k}}(B\rightarrow \infty)
  e^{it(\varepsilon_{\bm{k}^\prime}-\varepsilon_{\bm{k})}}:\bigl(S^+s_{\bm{k}^\prime\bm{k}}^-
  +S^-s_{\bm{k}^\prime\bm{k}}^+\bigr):\ .$ Up to neglected normal
  ordered terms of $\mathcal{O}(J^2)$, the operator $S^z(t)$ readily
  follows from integrating Eq.~(\ref{ansatzt}) with the initial
  condition $S^z(B\rightarrow \infty,t)$.  Using these steps, the
  formal result for the impurity magnetization reads
  \begin{equation}
    \langle S^z(t)\rangle=\frac{\tilde{h}}{2}+\sum_{\bm{k}\bm{k}^\prime} 
    \frac{\tilde{\gamma}_{\bm{k}\bm{k^\prime}}^2}{2}\biggl( e^{it(\epsilon_{\bm{k}}-\epsilon_{\bm{k^\prime}})} -\frac{1}{2}\biggr)n(\bm{k^\prime})[1-n(\bm{k})] \ .
  \end{equation}
  By solving the flow equations for the couplings
  $\tilde{\gamma}_{\bm{k}\bm{k^\prime}}$ for momenta close to the
  Fermi surface, it is therefore possible to directly obtain the
  long-time dynamics of the impurity magnetization. More details of
  the calculation can be found in Refs.~\cite{hackl09a,hackl09b}.  For
  isotropic couplings and using a flat band with the dimensionless
  range $[-1,1]$, this yields the asymptotic behavior
  \begin{equation}
    \langle S^z(t) \rangle=\frac{1}{2}\biggl( \frac{1}{\ln(t)-\frac{1}{\rho J}}+1+\rho J + \mathcal{O}(J^2)\biggr) \ .
  \end{equation}
  This asymptotic behavior is confirmed by numerical calculations
  depicted in Fig.~\ref{fig:2}.  For anisotropic couplings, the calculation is
  completely analogous and yields the result
  \begin{equation}
    \langle S^z(t) \rangle=\frac{1}{2}\biggl(1-\frac{\alpha^2}{2\tilde{j}^\parallel}t^{2\tilde{j}^\parallel} +\frac{\alpha^2}{2\tilde{j}^{\parallel}}\biggr) +\mathcal{O}(J^2) \ ,
  \end{equation}
  with the dimensionless coupling $\tilde{j}^{\parallel}=\rho\sqrt{J^{\parallel2}
    -J^{\perp2}}$ and $\alpha \approx \rho J^{\parallel}$ for $|J^\perp/J^\parallel
  |\geq 2$.  These results allow for two interesting observations: (i)
  the time-dependent contribution to $\langle S^z(t) \rangle$ is
  described by $J^{\parallel}(\Lambda=1/t)$, where $J^{\parallel}(\Lambda)$ flows
  according to the scaling equations. (ii) The asymptotic result
  $\langle S^z(t \rightarrow \infty)\rangle_{\Psi_i}-1/2=2(\langle S^z
  \rangle_{eq}-1/2)$ is obeyed.  Therefore, a factor of two
  distinguishes the asymptotic results for the equilibrium
  magnetization vs. the nonequilibrium magnetization.

  In this section we discussed a rare example where the nonequilibrium
  dynamics of an interacting many-particle system can be described
  analytically. The versatility of the flow equation approach to
  describe both equilibrium and nonequilibrium quantities with common
  scaling transformations allows to directly relate nonequilibrium
  relaxation laws to equilibrium properties obtained by a conventional
  renormalization group calculation. In our example, we could
  establish that nonequilibrium dynamics is set by the flowing
  coupling constants, evaluated at the energy scale equal to the
  inverse time scale of the relaxation process.  Furthermore, we could
  show that an equilibrium and a corresponding nonequilibrium
  observable in a steady state differ by a factor of two. This
  observation appears in a much wider class of problems, e.g. in a
  class of discrete models, where this result can be stated in form
  of a theorem \cite{moeckel09}.  The same factor two also appears for
  the real-time evolution in the weak coupling phase of the quantum
  Sine-Gordon model \cite{Sabio}, where it connects the equilibrium
  with the nonequilibrium mode occupation numbers.  It will be one of
  the central features for the real time evolution of the Hubbard
  model in Sec.~\ref{sec:floweqHubbard} since it defines the
  prethermalization and ensuing thermalization behavior.

  \section{Small interaction quenches in the Hubbard model}
  \label{sec:floweqHubbard}

  The forward-backward transformation scheme as it is depicted in
  Fig.~\ref{fig:1} became a fruitful tool also for examining the
  nonequilibrium properties of closed translation-invariant
  interacting quantum many-body systems.  Since the trace of the
  squared density operator $\rho$ does not change under unitary time
  evolution, a closed quantum system which has been initialized in a
  pure state (defined by $\text{Tr} [\rho^2]=1$) will never relax to a
  thermal state (defined by $\text{Tr} [\rho^2]<1$).  Nonetheless, the
  question of relaxation and thermalization can be addressed for
  expectation values of observables. As the conventionally studied
  simple observables typically only probe one- or two-particle aspects
  of the many-particle quantum state, there is the generic expectation
  that the full unitary evolution cannot be resolved; then the
  apparent dynamics of these observables becomes arbitrary close to
  that of a thermal state. This is, however, not the case for
  integrable models where the dynamics is limited by additional
  conserved integrals of motion. In the following we will show that
  the Hubbard model in more than one dimension, which is commonly
  considered nonintegrable, exhibits the phenomenon of
  prethermalization. It is characterized by the emergence of different
  relaxation times for various observables and has previously been
  observed numerically in cosmological models \cite{BergesPRL}.
This implies that there exists a transient time regime during which the expectation values of some observables have already relaxed to their long-time values while others have not.  The relaxation of the latter ones, however, is only delayed to a later time scale. 

  We start with imposing nonequilibrium initial conditions by a
  \emph{quantum quench}, i.e. a \emph{sudden} change in a parameter of
  the Hamiltonian: We start in the ground state of a noninteracting
  Fermi gas and assume that the strength of the Hubbard interaction
  remains small such that we always remain within the Fermi liquid
  phase of the model. Therefore we first test the nonequilibrium
  properties of a Fermi liquid beyond Landau's theory. Quenches
  outside of the Fermi liquid phase are discussed using the DMFT
  approach in Sec.~\ref{dmftquench}. Here we apply the
  forward-backward scheme which allows for an analytical \emph{ab
    initio} real-time analysis of the subsequent relaxation dynamics
  of the kinetic energy and the momentum distribution
  \cite{moeckel08,Moeckel2009PhD}.  It will turn out that the Fermi
  liquid picture provides a most suitable framework to understand the
  origin of prethermalization on the grounds of an analytical
  calculation. In the following paragraphs we will explain how
  prethermalization emerges from the interplay of the Pauli principle
  with quantum correlations induced by the two-particle interaction.

  The time evolution of a closed many-particle quantum system is fully
  described by its Hamiltonian. Let us assume that the latter contains
  different interactions between the various degrees of freedom on
  well-separated energy scales, for instance a two-particle
  interaction between electrons, an additional electron-phonon
  coupling or one to an external light field. Then one trivially
  expects that the relaxation of a generic excited state passes
  through a sequence of transient time regimes: each of them
  corresponds to an energy scale of the Hamiltonian and exhibits its
  own characteristic signatures; only afterwards full thermalization
  of all degrees of freedom may be reached.

  \emph{Prethermalization} of a Fermi liquid describes a similar
  sequence of time regimes which originates, however, from a single
  two-particle interaction term in the Hamiltonian. Moreover, it is
  reminiscent of a common effective approach for weakly interacting
  systems which discusses the Hamiltonian time evolution in terms of
  two different relaxation mechanisms: \emph{dephasing} of the initial
  state and two-particle \emph{scattering} between unrenormalized
  momentum modes. This corresponds to two distinct relaxation times:
  short time inelastic energy relaxation and long time elastic
  momentum relaxation. Prethermalization implies that these time
  scales separate and are observable in the relaxation behavior of
  different observables: 'bulk' quantities like the total kinetic or
  potential energy of the system are 'momentum mode averaged'
  quantities which already relax due to dephasing. This can be seen as
  an implication of energy-time uncertainty which allows for rapid
  \emph{energy exchange} at short times.  However, in a
  translationally invariant system, (quasi-) momentum is conserved and
  the momentum distribution can only relax by \emph{momentum exchange}
  due to scattering processes. Yet in a zero temperature Fermi liquid,
  scattering processes at the Fermi energy are suppressed by phase
  space restrictions which are imposed by the Pauli principle. Hence
  the momentum distribution represents the simplest example of a
  'momentum mode quantity' which relaxes on a later time scale. The
  separation of both time scales opens the transient regime of
  prethermalization during which a quasi-equilibrium description based
  on temperature cannot be defined for momentum mode quantities.

  We develop this scenario starting from a noninteracting Fermi gas at
  zero temperature, i.e. in the noninteracting ground state
  $\ket{\Omega_0}$. Its momentum distribution exhibits a discontinuity
  of size $Z^0=1$ at the Fermi energy. Then the two-particle Hubbard
  interaction is switched on instantaneously,
  \begin{equation}
    H = -
    \sum_{\langle i, j \rangle,\sigma} t_{ij} c^{\dagger}_{i \sigma}
    c_{j \sigma} + U \Theta(t) (n_{i,\up} - \frac 12)(n_{i \down}
    -\frac 12)
    \ ,
    \label{hubbardhamiltonian}
  \end{equation}
  where $c_i$ denotes a local fermionic destruction operator,
  $n_i=c^{\dagger}_ic_i$ the number operator in real space ($n_k$ in
  momentum space) and $\Theta(t)$ the Heaviside step function.  We
  assume that the nearest-neighbor hopping $t_{ij}$ is larger than the
  local on-site two-particle interaction $U \lesssim t_{ij} \equiv 1 $
  to permit a weak-coupling description.  The sudden change in the
  Hamiltonian initializes the system in a highly excited state and
  violates the fundamental prerequisite of Landau's equilibrium theory
  of a Fermi liquid, namely the adiabatic continuity of the
  noninteracting and the interacting Fermi system. Nonetheless we
  observe that important aspects of Landau's theory, in particular the
  quasiparticle picture of elementary excitations, are retained during
  the subsequent nonequilibrium dynamics.  Applying the unitary
  perturbation scheme of Fig. \ref{fig:1} to the creation operator in
  momentum space implements its Heisenberg equation of motion.
  Thereby, it exhibits the decay of a physical fermion $c_k^{\dagger}$
  into a superposition of various many-particle excitations which
  represent the 'dressing' of the noninteracting particle with
  particle-hole pairs due to interaction effects,
  \begin{equation}
    c_{k\sigma}(t=0,B) \stackrel{\rm time}{\rightarrow}  c_{k\sigma}(t,B) = h_{k \sigma}(t,B) \ c_{k\sigma} + M_{pqr\sigma \bar{\sigma}\bar{\sigma}}(t,B) \ c_{p\sigma}^{\dagger} c_{q\bar{\sigma}}^{\dagger} c_{r\bar{\sigma}} + \ldots
  \ .
  \end{equation}
  This ansatz is analog to the ansatz for the impurity magnetization
  (\ref{ansatzb}) in the ferromagnetic Kondo model. It transfers the
  time evolution of operators (and/or their flow under infinitesimal
  unitary transformations) to a set of coupled differential flow
  equations for the time and $B$-dependent prefactors $h_{k
    \sigma}(t,B)$ and $M_{pqr\sigma \bar{\sigma}\bar{\sigma}}(t,B)$.
  Higher order terms are truncated in a systematic way: only terms are
  kept which are relevant for a second order in $U$ result of the
  momentum distribution function $N_k(t) = \bra{\Omega_0} n_k
  (t)\ket{\Omega_0}$. At the Fermi surface, the coefficient $h_{k_F
    \sigma}(t) = \sqrt{Z(t)}$ relates to the time dependent value of
  the quasiparticle residue $Z$. The later mirrors the discontinuity
  of the momentum distribution function at the Fermi energy which is
  reduced under the time evolution. It approaches -- in a formal
  long-time limit -- a nonvanishing value $Z^{\rm NEQ}$ which
  mismatches the corresponding equilibrium value $Z^{\rm EQU}$ by a
  factor $\mu = \lim_{t \gg 1/U} (Z^0-Z^{\rm NEQ}(t))/(Z^0-Z^{\rm
    EQU})=2$.  A comparison with exactly solvable models,
e.g. the sudden squeezing of a harmonic oscillator \cite{moeckel09}, indicates
  that the mismatch is a nonperturbative effect of nonequilibrium
  dynamics, while its numerical value $\mu =2$ is a perturbative
  result in the weak-coupling limit. Such a mismatch is the generic
  behavior of many systems and observables \cite{moeckel09}. We have
  found it also comparing the nonequilibrium and equilibrium
  magnetization in the ferromagnetic Kondo model (cf.
  Sec.~\ref{sec:3}). For the Fermi liquid, the persistence of a
  nonvanishing quasiparticle residue up to late times has been
  confirmed recently \cite{Uhrig2009}.  Hence, up to a factor 
the momentum distribution still resembles that of a
  zero temperature Fermi liquid in equilibrium; therefore we conclude
  that a quasiparticle picture remains applicable. Note that, due to
  the mismatch, the corresponding quasiparticle momentum distribution
  is a nonequilibrium one: deviations from a description in terms of
  Landau quasiparticles are second order in $U$ and open phase space
  for additional quasiparticle scattering. Moreover, the kinetic
  energy $E_{\rm kin}(t) = \int d\epsilon_k \epsilon_k
  N_{\epsilon_k}(t)$ already relaxes to its final value during this
  first episode of the dynamics, which is on a time scale given by
  $t_1\sim 1/\rho_F U^2$. More details of the calculation can be found
  in Refs.~\cite{moeckel08,moeckel09,Moeckel2009PhD}.

  Corrections to the second order result imply a second stage of the
  dynamics. Motivated by the explicit calculation of one (out of many)
  forth order term we describe them by an effective kinetic evolution
  of the nonequilibrium momentum distribution using a quantum
  Boltzmann equation. The later describes the redistribution of
  occupation among quasiparticle momentum modes due to two-particle
  scattering events,
  \begin{equation}
    \frac{d N^{\rm QP}_k(t)}{dt} = \mathcal{I}_k [N^{\rm QP}(t)] = -U^2 \sum_{pqr} \mathcal{P}_{kpqr}[N^{\rm QP}(t)]  \delta^{k+p}_{q+r} \delta(\epsilon_k+\epsilon_p-\epsilon_q-\epsilon_r)
    \ .
  \end{equation}
  The scattering integral $\mathcal{I}_k [N^{\rm QP}(t)] $ contains
  the characteristic fermionic phase space factor which implements the
  constraints due to the Pauli exclusion principle on two-particle
  scattering,
  \begin{equation}
    \mathcal{P}_{kpqr}[N^{\rm QP}] =  \left[ N_k^{\rm QP} N_p^{\rm QP} (1-N_q^{\rm QP}) (1-N_r^{\rm QP}) - (1-N_k^{\rm QP})(1-N_p^{\rm QP}) N_q^{\rm QP} N_r^{\rm QP} \right]
    \ .
  \end{equation} 
  Note that the scattering integral vanishes for arbitrary Fermi-Dirac
  distributions including that one of the zero temperature Fermi gas
  $N^0$. Hence a thermal state is an attractive fixed point of the
  Boltzmann dynamics.  Since a quasiparticle picture has been
  established during the first stage we apply the quasiparticle
  momentum distribution of the transient state $N^{\rm QP}(t_1)$ as
  the initial condition of the further quantum Boltzmann dynamics.
  Linearizing the scattering integral around $N^0$ and noting that the
  displacement $\Delta N_k^{\rm QP}=N^{\rm QP}(t_1)-N^0$ is
  proportional to $U^2$ shows that this subsequent relaxation of the
  momentum distribution happens on a second time scale given by $t_2 =
  1/\rho^3 U^4$. For small interaction strength $t_2 \gg t_1$. This
  delayed relaxation behavior is reminiscent of the long
  nonequilibrium relaxation times in glasses. Those are explained by a
  ragged potential landscape: local minima represent transient states
  which are well-separated from the global thermodynamic ground state
  by energy barriers.  The example of the Fermi liquid shows that, in
  many-body quantum systems, analogue bottlenecks to the relaxation can
  be imposed by particle correlations. For Fermi systems at very low
  temperatures this is provided by the strong quantum statistical
  correlations due to the Pauli principle. However, this restriction
  is not exact, and in nonequilibrium significant quasiparticle
  scattering remains effective. Therefore thermalization of the
  momentum distribution can be expected on times $t>t_2$.  The
  existence of the prethermalization regime in the Hubbard model and
  its subsequent relaxation has been confirmed numerically in DMFT
  calculations, see Sec.~\ref{dmftquench}.

  Prethermalization is particularly relevant in ultracold Fermi gases:
  There it limits the efficiency of evaporative cooling since
  equilibration of the remaining atoms is very slow.  However,
  successive work has shown that this delayed thermalization may be
  turned into a feature and might simplify the observation of
  characteristic nonequilibrium physics at zero temperature, for
  instance nonequilibrium BCS behavior \cite{MoeckelNJP2010}.

  \begin{figure}
    \begin{center}
      \resizebox{0.65\columnwidth}{!}{
        \includegraphics{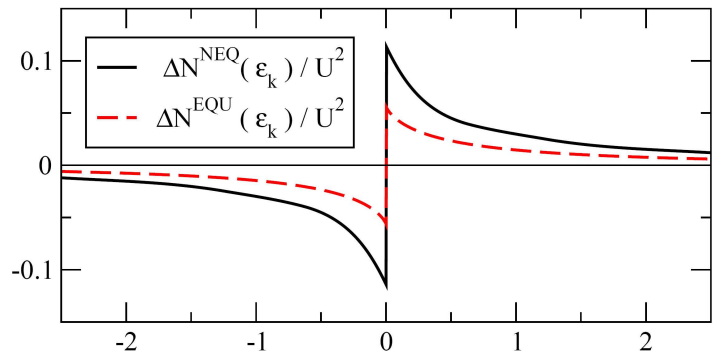}
      }
      \caption{Comparison of correlation-induced corrections to the
        momentum distribution function in equilibrium (broken line)
        and nonequilibrium (full line). One easily reads off the twice
        as large reduction of the quasiparticle residue at the Fermi
        energy in the nonequilibrium case as compared to equilibrium.}
      \label{fig:3} 
    \end{center}      
  \end{figure}

  \section{Transport beyond the linear response regime through Kondo dots}
  \label{sec:Kondotransport}

  A second important class of nonequilibrium problems is posed by
  transport beyond the linear response regime. While transport in the
  linear response regime essentially only probes the equilibrium
  ground state since the transport coefficients can be related to
  fluctuations in equilibrium, a correct description of the nonlinear
  response regime requires an understanding of the feedback of
  transport on the steady current-carrying state itself. In
  particular, many energy scales contribute to transport in this
  regime, which makes it conceptually difficult to access using
  traditional scaling techniques which focus on low-energy properties.

  An experimental and theoretical paradigm for transport beyond the
  linear response regime is realized in Coulomb blockade quantum dots
  in the Kondo regime, where a voltage bias~$V$ exceeding the Kondo
  energy scale~$T_{\rm K}$ can easily be applied
  \cite{GoldhaberGordon,Cronenwett,Schmid,vanderWiel}. The key
  theoretical insight for understanding this situation is the
  observation that the shot noise of the current across the dot leads
  to decoherence, which suppresses the coherent many-particle
  processes responsible for Kondo strong coupling physics
  \cite{Rosch2001}.

  \begin{figure}
    \begin{center}
      \resizebox{0.65\columnwidth}{!}{
        \includegraphics{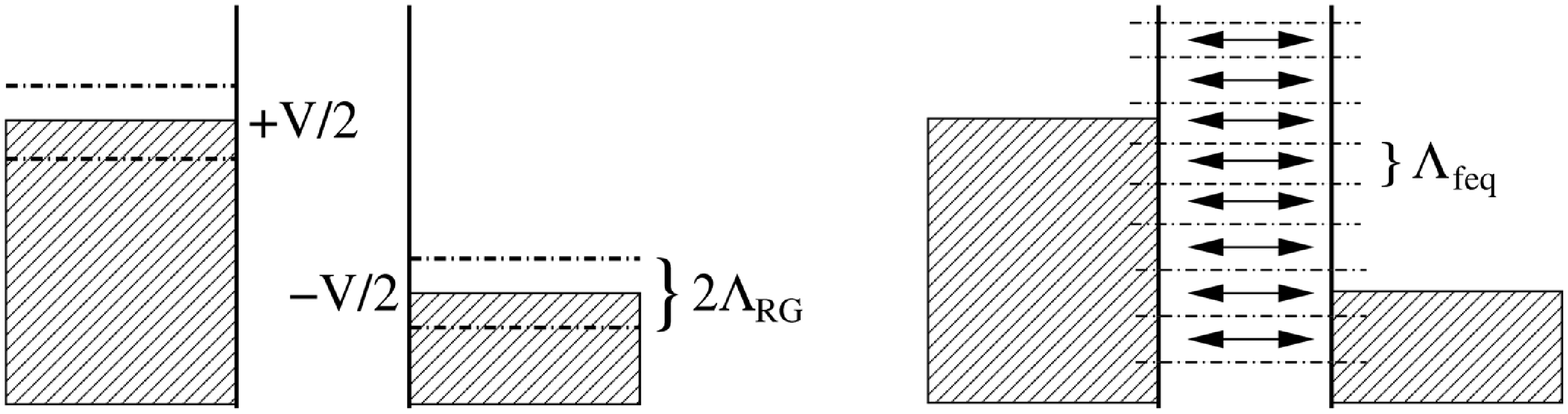}
      }
      \caption{Left: Conventional scaling picture where states are integrated
        out around the two Fermi surfaces with voltage bias~$V$ 
        (here depicted for cutoff~$\Lambda_{\rm RG}<V$).
        Right: Flow equation approach. Here all scattering processes with energy
        transfer $|\Delta E|\lesssim \Lambda_{\rm feq}$ are retained in 
        $H(\Lambda_{\rm feq})$.}
      \label{fig:compscaling} 
    \end{center}      
  \end{figure}

  The flow equation method captures this physics easily since it makes
  the Hamiltonian successively band-diagonal as opposed to projecting
  on the low-energy subspace as in traditional scaling approaches.
  This difference is depicted in Fig.~\ref{fig:compscaling} for
  transport between two leads at different chemical potentials.
  Clearly, all current-carrying states continue to contribute in the
  flow equation scheme.  It should be emphasized that there are other
  generalizations of conventional scaling approaches that also
  incorporate these processes
  \cite{Rosch2003,Rosch2005,Schoeller2000,Schoeller2009}.

  Explicitly, the Kondo Hamiltonian (\ref{hkondo}) for two leads takes
  the following form
  \begin{equation}
    H=\sum_{a,p,\alpha} (\epsilon^\pdag_p-\mu^\pdag_a)
    c^\dag_{ap\alpha} c^\pdag_{ap\alpha}\!
    +\sum_{a',a} J^\pdag_{a'a} \sum_{p',p} \vect{S}\, \cdot\, \vect{s}_{(a'p')(ap)} \ .
    \label{Kondo_nonequ}
  \end{equation}
  Here $a',a=l,r$ label the two leads and the chemical potentials are
  given by $\mu_{l,r}=\pm V/2$.  The conduction band electron spin
  operators are defined by
  $\vect{s}_{(a'p')(ap)}=\frac{1}{2}\sum_{\alpha,\beta}
  c^\dag_{a'p'\alpha} \vect{\sigma}^\pdag_{\alpha\beta}
  c^\pdag_{ap\beta}$, where $\vect{\sigma}$ are the Pauli matrices.
  The couplings $J_{a'a}$ describe the antiferromagnetic exchange
  interaction with the localized spin degree of freedom and are
  related by $J_{lr}^2=J^\pdag_{ll} J^\pdag_{rr}$ and
  $J_{ll}/J_{rr}=\Gamma_l/\Gamma_r$ if the model can be derived from
  an underlying Anderson impurity model.  $r\stackrel{\rm
    def}{=}\Gamma_l/\Gamma_r$ is the asymmetry parameter of the model.

\begin{figure}
    \begin{center}
      \resizebox{0.65\columnwidth}{!}{
        \includegraphics{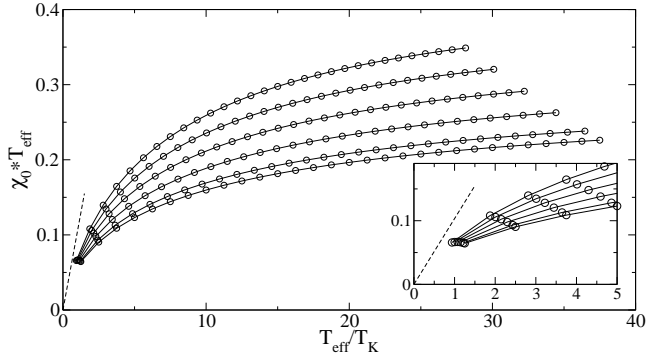}
      }
      \caption{Flow equation results for the static spin
        susceptibility $\chi_{0}$ for nonzero voltage bias for various
        asymmetry parameters~$r$ (increasing from bottom to top:
        $r=1.0, 1.4, 1.8, 2.2, 2.6, 3.0$).  The data is plotted as a
        function of the effective temperature
        $T_{\text{eff}}=V/(1+r)(1+r^{-1})$.  The dashed line is an
        exact result for the behavior in the $T_{\rm eff}\rightarrow
        0$ limit independent of asymmetry \cite{Fritsch2008}.}
      \label{fig:chi0rTeff} 
    \end{center}      
  \end{figure}

  The flow equation approach proceeds by using the canonical generator (\ref{defeta})
  and calculating the Hamiltonian flow (\ref{flowH})
  consistently including terms in third order of the coupling constant 
  (details can be found in Refs.~\cite{Kehrein2005,Fritsch2008,Fritsch2009}).
  In equilibrium this
  amounts to a two loop calculation and one recovers the well-known two loop scaling
  equation for the coupling constant $g=\rho\,J$ at the Fermi surface
  \begin{equation}
    \frac{dg}{d\ln\Lambda_{\rm feq}}=-g^2+\frac{g^3}{2}+O(g^4) 
    \ .
  \end{equation}
  The calculation is actually not different in the nonequilibrium case
  with voltage bias~$V$ except that now normal ordering is done with
  respect to the shifted Fermi seas (see Fig.~\ref{fig:compscaling}).
  This changes the scaling equations significantly for $\Lambda_{\rm
    feq}\lesssim \Gamma_{\rm rel}$ with
  \begin{equation}
    \Gamma_{\rm rel}=\sqrt{2\pi}\,\frac{V}{\ln^2(V/T_{\rm K})}\,\frac{1}{(1+r)(1+r^{-1})} \ ,
  \end{equation}
  which is just proportional to the shot noise generated by the
  current.  E.g. the scaling equation for the coupling $g_l$ at the
  Fermi surface of the left lead takes the form
  \begin{equation}
    \frac{dg_l}{d\ln\Lambda_{\rm feq}}=g_l\left(-\frac{g_l}{1+r^{-1}}
      +\frac{1}{2}\,\frac{\Gamma_{\rm rel}}{\Lambda_{\rm feq}+\Gamma_{\rm rel}}\right) 
      \ .
  \end{equation}
  One notices that if the coupling is not already too large at the
  scale $\Gamma_{\rm rel}$, then the strong coupling flow crosses over
  into a weak coupling flow with $g_l\rightarrow 0$ for $\Lambda_{\rm
    feq}\rightarrow 0$. This permits the controlled evaluation of
  physical quantities like dynamical spin susceptibility and
  $T$-matrix (for details see \cite{Fritsch2008,Fritsch2009}).
  Fig.~\ref{fig:chi0rTeff} for example shows the static spin
  susceptibility as a function of voltage bias, which crosses over
  quite accurately into the strong coupling equilibrium Bethe ansatz
  result in the limit $V\rightarrow 0$ (see inset of
  Fig.~\ref{fig:chi0rTeff}).  The generalization to the case with
  nonvanishing magnetic field can be found in Ref.~\cite{Fritsch2009},
  which leads to a more intricate interplay of different decoherence
  scales.

  \section{Nonequilibrium dynamical mean-field theory}
  \label{sec-dmft}

  Dynamical mean-field theory (DMFT) \cite{RMPGeorges} usually gives a
  reliable description of correlated systems in equilibrium whenever
  their physical properties are determined by local temporal
  fluctuations, and spatial correlations are not too important. Within
  DMFT, local correlation functions of the lattice model are obtained
  from an effective impurity problem which contains a single site of
  the lattice that is coupled to a bath of noninteracting degrees of
  freedom. This mapping becomes exact in the limit of infinite
  dimensions \cite{MetznerVollhardt}, and it can be formulated for both
  equilibrium and nonequilibrium situations
  \cite{Schmidt2002a}, using either the imaginary-time Matsubara or
  the Keldysh formalism. Nonequilibrium DMFT has been used to study
  nonlinear transport in the Falicov-Kimball model
  \cite{Freericks2006a,Freericks2008b,Joura2008a,Tsuji2008a,Tsuji2009a},
  as well as the behavior of the Falicov-Kimball model
  \cite{Eckstein2008a,Eckstein2009c} and the Hubbard model
  \cite{Eckstein2009a} when the interaction is changed abruptly, or
  slowly, as a function of time.  Furthermore,
  the method can be used for the description of time-resolved
  photoemission \cite{Freericks2009a,Eckstein2008c} and optical
  spectroscopy \cite{Eckstein2008b} in correlated systems.

  We now briefly review the nonequilibrium DMFT formalism.  A more
  detailed discussion of the technical aspects of this approach can be
  found in Refs.~\cite{Freericks2006a} and \cite{Eckstein2009b}. We
  then present the derivation of a class of closed-form
  self-consistency equations for the nonequilibrium case. The simplest
  self-consistency equation results for the semi-elliptic density of
  states, and was used already in several studies of interaction
  quenches in the Falicov-Kimball \cite{Eckstein2008a,Eckstein2009c}
  model and the Hubbard model \cite{Eckstein2009a}
  (cf.~Sec.~\ref{dmftquench}).
  
  The impurity problem of nonequilibrium DMFT is defined via the
  single-site action
  \begin{eqnarray}
    \label{action}
    \CS 
    = 
    -i\intC \!dt \,H_{\text{loc}}(t) 
    -i 
    \sum_{\sigma}
    \intC \!dt \intC \!dt'\,
    c^\mydag_{\sigma}(t)
    \Lambda(t,t') 
    c_{\sigma}(t')
  \end{eqnarray}
  on the Keldysh-contour $\CC$ that runs from $\tmin$ to some time
  $\tmax$ (i.e., the largest time of interest) on the real time axis,
  back to $\tmin$, and finally to $-i\beta$ along the imaginary time
  axis \cite{keldyshintro,Eckstein2009b}.  The first term on the
  right-hand side of this equation contains the dynamics due to the
  local Hamiltonian, e.g.,
  $H_{\text{loc}}(t)=U(t)n_{\uparrow}n_{\downarrow} -\mu
  (n_{\uparrow}+n_{\downarrow})$ in case of the Hubbard model with
  time-dependent interaction, e.g., as in
  Eq.~(\ref{hubbardhamiltonian}). The second term describes the
  hybridization of the site with an environment that replaces the rest
  of the lattice, and the hybridization function $\Lambda(t,t')$ must
  be determined self-consistently. From the action (\ref{action}),
  local contour-ordered correlation functions are obtained by
  computing the trace $\expval{A(t)B(t')\cdots}=\TR[\TC
  \exp(\CS)A(t)B(t)\cdots]/Z$, where $\TC$ is the contour-ordering
  operator, and real-time correlation functions can be read off from
  contour-ordered correlation functions by choosing the time-arguments
  appropriately. For the Hubbard model, the evaluation of those
  impurity correlation functions is the most demanding part of the
  DMFT solution. Real-time quantum Monte Carlo methods
  \cite{Werner2009a} were used successfully \cite{Eckstein2009a}
  (cf.~Sec.~\ref{dmftquench}), although the maximum accessible time is
  limited by the dynamical sign problem.  On the other hand, most
  nonequilibrium DMFT investigations have so far been performed for
  the Falicov-Kimball model, where Monte Carlo methods are not needed
  because one can derive a closed set of equations of motion for the
  impurity Green functions \cite{Brandt1989a}, which is then solved on
  the real time axis.

  To determine the hybridization function $\Lambda(t,t')$ self-consistently,  
  one must first compute the local self-energy $\Sigma$ from the Dyson equation 
  of the impurity model,
  \begin{equation}
    \label{dyson-imp}
    G=[i\partial_t + \mu -\Lambda - \Sigma ]^{-1},
  \end{equation}
  where $G(t,t')=-i\expval{\TC c_\sigma(t) c_\sigma^\dagger(t')}$ is
  the local Green function. Here and in the following, correlation
  functions are matrices in their contour-time
  variables. Matrix multiplication denotes a convolution along the
  contour $\CC$, the identity is the contour delta-function
  \cite{keldyshintro}, and the operator $\partial_t$ denotes the
  time derivative. Furthermore, we restrict ourselves to the
  paramagnetic phase and omit spin indices. Because there is usually
  no translational invariance in time in nonequilibrium situations,
  Eq.~(\ref{dyson-imp}) must be solved in the time domain instead of
  the frequency domain.  Next, the momentum-dependent Green function
  $G_\Vk=-i\expval{\TC c_\Ks(t) c_\Ks^\dagger(t')}$ of the lattice
  model is obtained from the lattice Dyson equation
  \begin{equation}
    \label{dyson-latt}
    G_\Vk=[i\partial_t + \mu -\epsilon_\Vk - \Sigma ]^{-1},
  \end{equation}
  where spatial homogeneity has been assumed, and $\epsilon_\Vk$ are
  the band energies. The DMFT self-consistency is finally closed by
  computing the local Green function at a given site $j$ in the
  lattice model, $G_{j\sigma} = \sum_\Vk |\expval{j|\Vk}|^2 G_\Ks$,
  and requiring it to be equal to the local Green function $G$ of the
  impurity model. If the band energies $\epsilon_\Vk$ are
  time-independent, i.e., when there is no external electrical field,
  the momentum summation can be reduced into a single energy integral
  \begin{eqnarray}
    \label{gloc}
    G = \int\!d\epsilon\,\rho(\epsilon) G(\epsilon),
  \end{eqnarray}
  where $\rho(\epsilon)=\sum_\Vk |\expval{j|\Vk}|^2
  \delta(\epsilon-\epsilon_\Vk)$ is the local density of states, and
  $G(\epsilon_\Vk)=G_\Vk$ is the momentum-dependent Green function that
  depends on momentum only via the band energy $\epsilon_\Vk$.

  The numerical solution of Eqs.~(\ref{dyson-imp}) and (\ref{dyson-latt}) is 
  achieved either via an explicit matrix inversion  \cite{Freericks2008b} (after 
  the contour $\CC$ is discretized into $N$ time slices, and all contour 
  Green functions, the derivative operator, and the identity are transformed 
  into $N$-dimensional matrices), or the equations are rewritten as a set 
  of integro-differential equations of Volterra type which are then solved 
  by standard numerical algorithms \cite{Eckstein2009b}. Although both 
  approaches are rather well behaved and can be carried on to quite long 
  times $\tmax$ (see, e.g., \cite{Eckstein2009c}) the numerical effort can 
  add up considerably if the $\Vk$ summation or $\epsilon$ integration in 
  Eq.~(\ref{gloc}) requires a large number of $\Vk$-points 
  \cite{Freericks2006a,Freericks2008b}. It is therefore desirable to find cases 
  in which Eqs.~(\ref{dyson-imp})-(\ref{gloc}) can be further simplified. Such 
  a simplification occurs for the semi-elliptic density of states (with
  bandwidth $W=4$), 
  \begin{equation}
    \label{rho-semi}
    \rho(\epsilon) = \frac{\sqrt{4 -\epsilon ^2}}{2 \pi},
  \end{equation}
  which corresponds to nearest-neighbor hopping on the Bethe lattice
  \cite{Economou1979a,Mahan2001a,Eckstein2005a,Kollar2005a}, or a
  particular kind of long-range hopping on the hypercubic lattice
  \cite{Bluemer2003a}.  If the density of states (\ref{rho-semi}) is
  inserted in Eq.~(\ref{gloc}), the self-energy can be eliminated from
  Eqs.~(\ref{dyson-imp})-(\ref{gloc}), such that one obtains a closed
  expression for the Weiss field \cite{Eckstein2008a},
  \begin{equation}
    \label{selfconsistency-bethe}
    \Lambda = G.
  \end{equation}
  This closed form of the self-consistency reduces the DMFT
  self-consistency cycle to a repeated solution of
  Eq.~(\ref{selfconsistency-bethe}) and the single-site problem for
  the local Green function $G$. In case of the interaction quench in
  the Falicov-Kimball model the existence of a closed self-consistency
  is crucial for the derivation of an analytical solution, thus
  providing the unique opportunity to study the limit of infinite
  times. We will now give a detailed derivation of this relation from
  a slightly more general statement, which in principle allows to
  obtain similar closed form self-consistency equations for densities of
  states other than Eq.~(\ref{rho-semi}).
  
  Suppose that a density of states $\rho(\epsilon)$ is given and that
  its Hilbert transform,
  \begin{equation}
    g(z) = \int d\epsilon \, \frac{\rho(\epsilon)}{z-\epsilon},
    \label{hilbert}
  \end{equation}
  for complex frequency $z$ satisfies the equation
  \begin{equation}
    \label{eq:app-f}
    zg=1+\sum_{n=1}^{\infty} f_n g^n.
  \end{equation}
  with an analytical function $F(g)=\sum_{n=1}^{\infty} f_n g^n$ with
  real coefficients $f_n$.  Note that the coefficients $f_n$ can be
  obtained order by order by first expanding Eq.~(\ref{hilbert}) for large
  $z$ and inverting the resulting moment expansion into a series of
  $z$ in powers of $g$.  For example, $F(g)=g^2$ for the semi-elliptic
  density of states~(\ref{rho-semi}).

  Consider now square matrices $Z$ and $G$ that are related by
  \begin{align}
     \label{app-gloc}
     G &= \int d\epsilon \, \rho(\epsilon) \, G(\epsilon)
     \\
     \label{app-dyson}
     G(\epsilon) &=  (Z-\epsilon)^{-1}.
  \end{align}
  Then, as we will show in the remainder of this section, the matrices
  $Z$ and $G$ are determined by the matrix equation
  \begin{equation}
    \label{eq:app-f-matrices}
    ZG=1+\sum_{n=1}^{\infty} f_n G^n.
  \end{equation}
  which is analogous to the scalar equation (\ref{eq:app-f}).  By
  setting $Z=i\partial_t + \mu -\Sigma$ we see that
  Eqs.~(\ref{app-gloc}) and (\ref{app-dyson}) correspond to
  Eqs.~(\ref{gloc}) and (\ref{dyson-latt}), respectively. Equation
  (\ref{eq:app-f}) can be transformed into Eq.~(\ref{dyson-imp}) if we
  multiply it with $G$ from the right and set
  \begin{equation}
    \label{thatisit}
    \Lambda  = \sum_{n=0}^{\infty} f_{n+1} G^{n}.
  \end{equation}
  This last equation (which reduces to
  Eq.~(\ref{selfconsistency-bethe}) for $F(g)=g^2$) provides the
  desired closed form of the self-consistency equation.  We note that if
  $F(g)$ is not a polynomial of finite degree, one can think of a
  suitable expansion in terms of orthogonal polynomials, e.g.,
  Chebyshev polynomials \cite{Weisse2006a}. Because orthogonal
  polynomial of $G$ (which involve $n$-fold convolutions) can be
  computed recursively, this might still be more efficient that
  performing the $\epsilon$ integral (\ref{gloc}) with one matrix
  inversion per integration point.

  To prove Eq.~(\ref{eq:app-f-matrices}) for general square matrices,
  we multiply the equation $(Z-\epsilon)G(\epsilon)$ $=$ $1$ with
  $\rho(\epsilon)=-{\rm Im}[g(\epsilon+i0)]/\pi$, and integrate over
  $\epsilon$. Using Eq.~(\ref{app-gloc}), this yields
  \begin{equation}
    Z G = 1 -\frac{1}{\pi}\int d\epsilon \,
    \epsilon\,{\rm Im}[g(\epsilon+i0)]\,G(\epsilon).
  \end{equation}
  Employing Eq.~(\ref{eq:app-f}) for the scalar $g$, one can replace
  $\epsilon \, {\rm Im}[g(\epsilon+i0)]$ by
  $\text{Im}[F(g(\epsilon+i0))]$, leading to
  \begin{equation}
    Z G = 1 -\frac{1}{\pi} \sum_n f_n \int d\epsilon \,
    {\rm Im}[g(\epsilon+i0)^n]
    \,G(\epsilon).
  \end{equation}
  It thus remains to prove that
  \begin{equation}
    G^n = -\frac{1}{\pi} \int d\epsilon \,
    {\rm Im}[g(\epsilon+i0)^n]G(\epsilon)
    \label{eq:app-gn}
  \end{equation}
  holds for any integer $n\geq1$, which is done by induction: The
  initial step ($n=1$) follows from the definition (\ref{app-gloc}).
  For the induction step, consider
  \begin{eqnarray}
    (G)^{n+1} 
    &=&
    (G)^{n} G
    \nonumber \\
    &\stackrel{(i)}{=}&
    \frac{1}{\pi^2} \int d\epsilon \,d\epsilon' \,
    {\rm Im}[g(\epsilon+i0)^n]{\rm Im}[g(\epsilon'+i0)] 
    \,G(\epsilon)  G(\epsilon')
    \nonumber \\
    &\stackrel{(ii)}{=}&
    \frac{1}{\pi^2} \int d\epsilon\, d\epsilon' \,
    {\rm Im}[g(\epsilon+i0)^n]{\rm Im}[g(\epsilon'+i0)] 
    \,G(\epsilon)  
    \left(\frac{Z-\epsilon+\epsilon'-Z}{\epsilon'-\epsilon +i0}\right)
    G(\epsilon')
    \nonumber \\
    &\stackrel{(iii)}{=}&
    \frac{1}{\pi^2} \int d\epsilon \,d\epsilon' \,
    \frac{{\rm Im}[g(\epsilon+i0)^n]{\rm Im}[g(\epsilon'+i0)]}{\epsilon'-\epsilon +i0} 
    \,(G(\epsilon) - G(\epsilon'))
    \nonumber \\
    &\stackrel{(iv)}{=}&
    -\frac{1}{\pi} \int d\epsilon \, G(\epsilon)\,
    (g(\epsilon+i0)^n{\rm Im}[g(\epsilon+i0)]
    +{\rm Im}[g(\epsilon+i0)^n]g(\epsilon-i0))
    \nonumber \\
    &=&
    -\frac{1}{\pi} \int d\epsilon \,{\rm Im}[g(\epsilon+i0)^{n+1}]\,G(\epsilon) \ .\nonumber
  \end{eqnarray}
  In step (i), proposition (\ref{eq:app-gn}) is used. In (ii), the
  term in braces is unity, and in (iii), we use
  Eq.~(\ref{app-dyson}).  To proceed to (iv) one performs one of the
  two energy integrals by making use of the spectral representation
  \begin{equation}
    g(z)^n = -\frac{1}{\pi}\int d\epsilon \frac{{\rm Im}[g(\epsilon+i0)^n]}{z-\epsilon}.
  \end{equation}  
  The latter holds because $g(z)^n$ is analytic in the upper half
  plane.  Summing up the terms in step (iv) and using
  $g(\epsilon-i0)=g(\epsilon+i0)^*$ completes the induction, and hence
  the proof of the closed self-consistency equation (\ref{thatisit}).

  \section{Interaction quenches in the Hubbard model in DMFT}
  \label{dmftquench}

  We will now discuss the relaxation dynamics of Hubbard models
  after a sudden change in the interaction parameter from $0$ to a
  finite value of $U$, see Eq.~(\ref{hubbardhamiltonian}).  Since
  only the dynamics of the interacting fermions is considered, but
  no coupling to an external bath is present, it is not
  immediately clear whether the system will thermalize, i.e.,
  whether after sufficiently long times it can be described by the
  thermal state that is predicted by equilibrium statistical
  mechanics.  Many models of isolated many-body systems have so
  far been investigated (see, e.g.,
  Refs.~\cite{Heims65,Girardeau1969,Sengupta2004a,Rigol2006,Cazalilla2006a,Rigol2007a,Kollath2007,Manmana07,Eckstein2008a,moeckel08,Kollar2008,Rigol2008a,Rossini2009b,Barmettler2009a,Eckstein2009a},
  and \cite{Dziarmaga2009} for a recent review), and in fact
  thermalization is only rarely observed.  In this section we will
  discuss results obtained with nonequilibrium DMFT, which show
  that the system can thermalize quickly for certain values of the
  interaction $U$ \cite{Eckstein2009a,Eckstein2009b}.

  \begin{figure}
    \centering
    \includegraphics[width=0.75\textwidth]{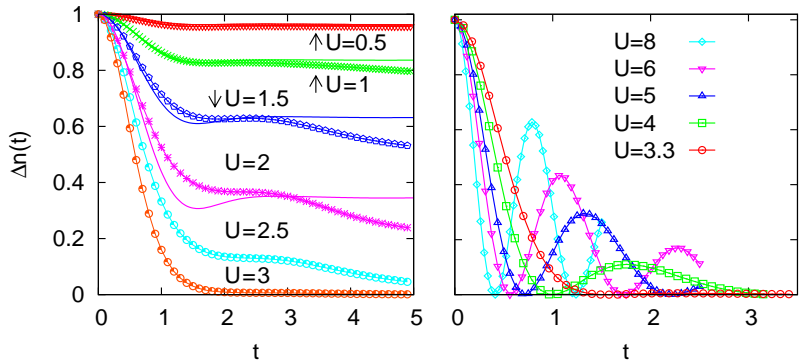}
    \caption{Fermi surface discontinuity $\Delta n$ after quenches
      from $U=0$ to $U\leq3$ (left panel) and $U\geq3.3$ (right
      panel) \cite{Eckstein2009a,Eckstein2009b}. The expected
      thermal value for this quantity is zero, corresponding to a
      continuous momentum distribution at finite temperature.  The
      quarter-bandwidth $V$ is set to unity.  Solid lines in the
      left panel indicate the flow equation prediction for the
      prethermalization plateau \cite{moeckel08} and the transient
      behavior \cite{moeckel09} for small values of
      $U$.\label{fig:hubb_dmft}}
  \end{figure}

  For a quench from the ground state at $U$ $=$ 0 to finite values of
  $U$ the DMFT equations for the paramagnetic phase
  (Sec.~\ref{sec-dmft}) were obtained in
  Refs.~\onlinecite{Eckstein2009a,Eckstein2009b}, using the
  semi-elliptic density of states~(\ref{rho-semi}) 
  and real-time quantum Monte Carlo methods \cite{Werner2009a} to
  obtain the Green function for the action (\ref{action}).
  Fig.~\ref{fig:hubb_dmft} shows the jump $\Delta n(t)$ in the
  momentum distribution at the Fermi surface as a function of
  time, in separate panels for small and large values of the final
  interaction parameter $U$.  Clearly the time evolution after an
  interaction quench in the Hubbard model depends sensitively on
  the parameter $U$. Note that $\Delta n(t)$ remains finite for a
  finite time after the quench; in the case of a local self-energy
  as in DMFT this is due to the relation of $\Delta n(t)$ to the
  retarded Green function at $\epsilon=0$~\cite{Eckstein2009a}.
  From this relation one can infer that the collapse of $\Delta n$
  is closely related to the decay of charge excitations that are
  created at the Fermi surface by the quench. It should be noted
  that as long as $\Delta n(t)$ is finite the system is not yet
  thermalized, because a finite jump is possible in a Fermi liquid
  in thermal equilibrium only at zero temperature, whereas the
  quenched system has finite excitation energy.

  For quenches to $U$ $\le$ $3$ the Fermi surface discontinuity
  $\Delta n(t)$ remains finite for times $t$ $\leq$ $5$ (left
  panel in Fig.~\ref{fig:hubb_dmft}).  The plateau in $\Delta
  n(t)$ at intermediate times is given by $2Z-1$, where $Z$ is the
  quasiparticle weight in equilibrium at zero temperature and
  interaction $U$ (cf. Sec.~\ref{sec:floweqHubbard}). On the
  other hand, the double occupation does essentially relax 
  to its thermal value on this timescale
  \cite{Eckstein2009a}, confirming that the potential energy (and
  therefore also the kinetic energy) relax quickly, whereas the
  occupation of individual states still changes.  Interestingly
  the prethermalization plateau remains well visible even for
  quenches to relatively large $U\lesssim2.5$, even though the
  timescales $V/U^2$ and $V^3/U^4$ are then no longer well
  separated.  Not only the prethermalization plateaus, but also
  the transient behavior predicted from the flow equation
  analysis~\cite{moeckel09} agrees well with the numerical DMFT
  results for $U\lesssim1.5$.  The prethermalization plateaus for
  small $U$ are due to the vicinity of the integrable point at
  $U=0$, as discussed further below.

  The relaxation dynamics show different characteristics for
  quenches to large $U$ (right panel of Fig.~\ref{fig:hubb_dmft}),
  namely so-called collapse-and-revival oscillations with
  approximate frequency $2\pi/U$. These are to due the vicinity of
  the atomic limit ($V$ $=$ $0$), for which the propagator
  $e^{-iHt}$ is periodic with period
  $2\pi/U$-periodic~\cite{Greiner}. For finite $V$ these
  oscillations are damped and decay on timescales of order $1/V$.
  For the double occupation these oscillations are not centered
  around the thermal value, but rather around a different value
  that can be explained using perturbation theory for strong
  coupling \cite{Eckstein2009a}. Somewhat analogously to the
  situation at small coupling the relaxation to the thermal state
  is thus delayed because the system is trapped in a metastable
  state close to an integrable point.

  Interestingly both the weak-coupling prethermalization plateau
  in $\Delta n(t)$ and the strong-coupling oscillations disappear
  in a small region of interaction parameters around
  $U_c^{\text{dyn}}\approx3.2$ \cite{Eckstein2009a}.  For quenches
  to values of $U$ near $U_c^{\text{dyn}}$ the system thermalizes
  very quickly. Not only the Fermi surface discontinuity and the
  double occupation relax to their thermal values, but in fact the
  retarded nonequilibrium Green function relaxes to the
  corresponding equilibrium quantity~\cite{Eckstein2009b}. It is
  therefore justified to say that the system indeed thermalizes in
  this case, because a large set of observables tend to the
  thermal value predicted by equilibrium statistical mechanics.
  Note that the statistical theory contains no adjustable
  parameter, because the effective temperature $T^*$ is determined
  by the energy of the system after the quench, for example
  $T^*=0.84$ for the quench to $U=3.3$.  The sharp crossover in
  the relaxation parameter is intriguing, not least because the
  relation to the equilibrium Mott metal-insulator transition is
  not obvious. The critical endpoint of the latter is located at
  $T_c \approx 0.055V$~\cite{RMPGeorges}, so that little of it is
  visible at the much higher temperatures $T^*$. For quenches of
  the anisotropy parameter in Heisenberg chains a remarkably
  similar behavior was found~\cite{Barmettler2009a}. There the
  staggered magnetization indeed relaxes fastest to zero for
  quenches to the equilibrium critical value.

  It follows from these results that an isolated fermionic
  many-body system can thermalize merely under the time evolution
  with the interacting Hamiltonian, without requiring coupling to
  a bath.  In the vicinity of an intermediate value
  $U_c^{\text{dyn}}$ the system quickly thermalizes, but for
  smaller or larger coupling thermalization has not been observed
  numerically for the short available observation times.

  \section{Interaction quenches in the $1/r$ Hubbard chain}
  \label{chainquench}

  For comparison we now discuss results for the one-dimensional
  $1/r$ Hubbard model, which was originally proposed and solved by
  Gebhard and Ruckenstein~\cite{Gebhard92}. Its hopping amplitudes
  are given by $t_{mj}$ $=$ $(-iW/2L)(-1)^{m-j}/\sin[\pi(m-j)/L]$
  with periodic boundary conditions, leading to a linear
  dispersion $\epsilon_k$=$Wk/(2\pi)$, where $W(=1)$ is the
  bandwidth.  For $U$ $\geq$ $-1$ this model can be mapped to an
  effective free Hamiltonian for hard-core
  bosons~\cite{Gebhard92,Gebhard94b} for which the spectrum can be
  determined at once.  For half-filling a Mott-Hubbard
  metal-insulator transition occurs at the interaction strength
  $U_c=1$ in this model, and the Mott gap is $\Delta=U-U_c$ in the
  insulating phase.

  Based on this solution, the exact time evolution was obtained in
  Ref.~\onlinecite{Kollar2008} for a quench from $U=0$ (or $U=\infty$)
  to a finite value of $U$. For the quench from $U=0$ to finite $U$ at
  half-filling the double occupation has the noninteracting value
  $\frac14$ at time $t=0$ and relaxes to a new constant value with
  algebraic decay,
  \begin{align}
    d(t)
    &=
    \frac{1}{8}
    -
    \frac{(1-U)^2}{16U}
    -
    \frac{(1-U^2)^2}{16U^2}
    \ln\Big|\frac{1-U}{1+U}\Big|
    -\frac{\cos(Ut)\cos(t)}{2Ut^2}    
    +
    O\Big(\frac{1}{t^3}\Big)
    \,.
  \end{align}
  In general the long-time limit of $d(t)$ does not agree with the
  thermal prediction $d_\text{therm}$.  For example, for quenches
  to the critical value $U_c=1$ the stationary value is
  $d(t=\infty)=0.125$, and this differs too from the thermal
  prediction $d_\text{therm}=0.098$ that is obtained from
  equilibrium results~\cite{Gebhard92} at the temperature that
  gives the same mean energy as the time-evolved state.

  The reason for the nonthermal steady state in this model lies in
  its integrability. The interacting fermionic Hamiltonian $H$ can
  be mapped on an effective Hamiltonian without interactions,
  i.e.,
  \begin{align}
    H = \sum_{\alpha}\epsilon_\alpha\;\calI_\alpha\,,
    \label{integrableH}
  \end{align}
  where the $\calI_\alpha$ have the eigenvalues 0 or 1 and commute
  with each other (and hence with $H$).  This is a rather strong case
  of integrability, because not only are there many constants of
  motion (their number is proportional to the system size), but also
  the energy levels can be populated independently (unlike, e.g., in many other
  models that can be solved by Bethe ansatz).  As a consequence the
  fundamental postulate of statistical mechanics is not fulfilled,
  which states that all accessible microstates are assumed to be equally
  probable in the ensemble description of the equilibrium state. For
  an integrable system~(\ref{integrableH}), however, all the
  $\calI_\alpha$ remain at their values in the initial state at $t=0$.
  That this restriction leads to nonthermal steady states has been
  observed in a variety of
  models~\cite{Girardeau1969,Sengupta2004a,Rigol2006,Cazalilla2006a,Rigol2007a,Manmana07,Eckstein2008a},
  and also in experiments with cold atomic
  gases~\cite{Kinoshita2006a}.

  \section{Statistical description of nonthermal steady states with
    generalized Gibbs ensembles}
  \label{gge}

  As described in the previous subsection, standard statistical
  mechanics cannot be expected to predict the steady state after a
  quench in an integrable system correctly. This leads to the
  question whether a suitably generalized statistical theory can
  make the correct prediction. The general procedure to construct
  the statistical operator in statistical mechanics is to maximize
  the entropy, and to take conserved quantities into account by
  fixing them on average using Lagrange multipliers
  \cite{Balian91a}. This means that for the integrable
  Hamiltonian~(\ref{integrableH}) all constants of motion
  $\calI_\alpha$ should be fixed on average such that they yield
  the correct initial expectation
  value~\cite{Rigol2006,Rigol2007a}, leading to a generalized
  Gibbs ensemble (GGE)
  \begin{align}%
    \rho_{\text{GGE}}
    &\propto
    e^{-\sum_\alpha\lambda_\alpha\calI_\alpha}
    \,,\label{eq:gge}
  \end{align}%
  with $\lambda_\alpha$ determined from
  $\expval{\calI_\alpha}_\text{GGE}=\expval{\calI_\alpha}_{0}$.
  For the $1/r$ Hubbard chain discussed in the previous
  subsection, the GGE prediction for the stationary value of the
  double occupation agrees precisely with the calculated long-time
  limit~\cite{Kollar2008}.  However, GGEs have been found to fail
  for some observables in other models~\cite{Rigol2006}.

  It is possible to formulate sufficient criteria for the steady
  state to be correctly described by the ensemble (\ref{eq:gge}).
  In general this depends on both the initial state
  $\ket{\psi(t=0)}$ and the observable $A$~\cite{Kollar2008}.  The
  long-time average of $\bra{\psi(t)}A\ket{\psi(t)}$ is given by
  \begin{align}%
    \overline{\expval{A}}
    &=
    \sum_{\bm{m}}
    \bra{\bm{m}}
    A
    \ket{\bm{m}}
    |\braket{\bm{m}}{\Psi(t=0)}|^2
    \,,\label{eq:average_diag}
  \end{align}%
  provided that the spectrum of $H$ is nondegenerate. Note that
  $\overline{\expval{A}}$ equals the long-time limit of
  $\expval{A}$, if the latter exists. Here the eigenbasis of the
  constants of motion, $\calI_\alpha\ket{\bm{m}}$ $=$
  $m_\alpha\ket{\bm{m}}$, was used. Let us assume that the
  constants of motion only have eigenvalue $m_\alpha=0,1$ and can
  be represented as bosons or fermions according to
  $\calI_\alpha=a_{\alpha}^\mydagger a_{\alpha}^{\phdagger}$
  (cf.~\cite{Kollar2008} for a more general case)). Consider then
  an observable $A$ that can be written as a linear combination of
  products of $n$ creation operators $a_{\alpha_i}^\mydagger$
  followed by $n$ annihilation operators $
  a_{\alpha_j}^{\phdagger}$.  The long-time average
  $\overline{\expval{A}}$ is given by a linear combination of
  expectation values
  $\langle{\prod_{i=1}^n\calI_{\alpha_i}}\rangle_{t=0}$ in the
  initial state; off-diagonal terms drop out due to the diagonal
  character of (\ref{eq:average_diag}). On the other hand the GGE
  expectation value decouples each constant of motion from the
  others and instead involves
  $\prod_{i=1}^n\langle{\calI_{\alpha_i}}\rangle_{t=0}$, where the
  condition on the parameters $\lambda_\alpha$ has been used. If
  the expectation value of the product equals the product of these
  expectation values for each $\alpha_i$ that occurs in the
  observable $A$, then the GGE is guaranteed to describe the
  expectation value of $A$ in the steady state correctly.
  
  We see that for the GGE prediction to be correct, the
  $\calI_{\alpha_i}$ that occur in $A$ must in a certain sense not be
  too correlated in the initial state, so that the expectation value of
  their product factorizes. Unfortunately the physical meaning of this
  condition is not obvious and certainly depends on the way in which
  the original degrees of freedom appear in the constants of motion.
  Nevertheless the GGE is certain to work for quadratic operators
  $a_{\alpha}^\mydagger a_{\beta}^{\phdagger}$, and this is the reason
  why the GGE succeeds for the double occupation in the $1/r$ Hubbard
  model, which can be expressed precisely as a sum of such quadratic
  operators~\cite{Gebhard94b}. 

  In any case we can conclude that statistical mechanics, when
  applied properly, does indeed predict the steady state of an
  integrable many-body system that is isolated and not coupled to
  any baths.  The fact that GGEs cannot describe all expectation
  values correctly is actually not surprising. Even standard
  statistical mechanics fails for the expectation values of
  specially crafted operators, such as powers of the Hamiltonian
  or projectors onto energy eigenstates, which remain at their
  initial values and thus never thermalize. But these observables
  are typically highly nonlocal and their expectation values
  correspond to correlation functions of very high order. Since
  they are essentially impossible to measure they are not of
  practical importance, implying no severe limitation on the
  applicability of equilibrium statistical mechanics.
  
  So far we have discussed integrable systems in the sense of
  Eq.~(\ref{integrableH}), i.e., systems of interacting particles
  whose Hamiltonian can be transformed into a basis in which the
  new effective degrees of freedom are noninteracting, and their
  occupation numbers are thus constants of motion. However, any
  noninteracting Hamiltonian is of course also integrable, e.g.,
  the Hubbard model with $U=0$, $H_0 = \sum_{\bm{k}\sigma}
  \varepsilon_{\bm{k}} c^{\mydagger}_{\bm{k}\sigma}
  c^{\phdagger}_{\bm{k}\sigma}$.  In this case the conserved
  momentum occupation numbers $c^{\mydagger}_{\bm{k}\sigma}
  c^{\phdagger}_{\bm{k}\sigma}$ play the role of the conserved
  $\calI_{\alpha}$, so that thermalization is impossible when
  quenching to $H_0$.

  Moreover, noninteracting Hamiltonians of course also provide
  useful starting points for interaction quenches, as used in
  Sec.~\ref{sec:floweqHubbard}.  As discussed there, a quench to
  a small value of the interaction parameter will lead to
  prethermalization on an intermediate time
  scale~\cite{moeckel08}, which is due the vicinity of the
  integrable point at $U=0$. In fact, the prethermalized
  expectation value of an observable $A$ can be obtained by using
  perturbation theory and taking the long-time limit (provided
  $[A,H_0]=0$)~\onlinecite{moeckel09}. It is therefore useful to
  view the prethermalization plateau as due to the conservation of
  certain perturbed constants of motion
  $\widetilde{\calI}_{\alpha}$, which hinder thermalization on
  intermediate time scales. It is then possible to construct a
  generalized Gibbs ensemble $\widetilde{\rho}_\text{GGE}$ based
  on these perturbatively conserved quantities, which predicts a
  stationary value that agrees precisely with the long-time limit
  in second order perturbation theory, provided certain
  factorization conditions are fulfilled~\cite{Kollar2010}.
  Namely, for an observable $A=\prod_i\calI_{\alpha_i}$ the
  prethermalization plateau, which occurs for small quenches away
  from $H_0$, is correctly predicted by
  $\widetilde{\rho}_\text{GGE}$ if
  $\expval{\prod_i\calI_{\alpha_i}}$ =
  $\prod_i\expval{\calI_{\alpha_i}}$. Here the expectation value
  is to be taken in the (perturbed) eigenstate after the quench.
  Again it is not surprising to see that only sufficiently simple
  observables can be expected to be correctly predicted by the
  statistical theory. In the simplest case, the GGE prediction for
  the prethermalization plateau is correct for the observables
  $\calI_{\alpha}$, e.g., the momentum distribution for quenches
  to small $U$ in the Hubbard model.  The agreement between the
  time-evolved state and the GGE prediction shows that
  prethermalization plateaus evolve continuously from the
  nonthermal steady state in integrable models, at least
  perturbatively.

  \section{Conclusion}

  For the investigation of the nonequilibrium behavior of
  correlated systems robust theoretical techniques are required,
  because both the interaction between particles and the time
  evolution for sufficiently long times must be described
  reliably. On the one hand, we discussed the flow equation
  approach and its applications to isolated many-body systems in
  nonequilibrium such as the ferromagnetic Kondo model and the
  Hubbard model at weak coupling, as well as nonlinear transport
  through Kondo dots. This method has the advantage that the
  emergence of time and energy scales is explicit and transparent.
  For example, for small interaction quenches in Hubbard models
  the method reveals that the system prethermalizes, i.e., that
  nonthermal states form on an intermediate time scale. 
Quantum Boltzmann dynamics then shows that these nonthermal states thermalize
on a longer time scale \cite{moeckel08,moeckel09}.
  On the other hand, we discussed applications of nonequilibrium
  dynamical-mean field theory, which maps a lattice system onto an
  effective single-site problem that can be solved numerically.
  The numerical results for interaction quenches in the Hubbard
  model confirm the prethermalization scenario and also show that
  thermalization can indeed occur on short timescales in an
  isolated many-body system at intermediate coupling.  We compared
  with results for special solvable models, in which nonthermal
  steady states occur due to the presence of many constants of
  motion.  Within second order perturbation theory, the statistical predictions of generalized Gibbs
  ensembles, which take these constants of motion into account,
  show that prethermalization in nonintegrable systems and
  nonthermal states in integrable systems can be related.

  \medskip
{\small
  We thank M. Rigol for useful discussions. Financial support by the
  SFB 484 and TTR 80 of the Deutsche Forschungsgemeinschaft (DFG) is
  gratefully acknowledged. A.H. acknowledges additional support through SFB 608 and
financial support through SFB/TR 12 of the DFG. S.K. and M.M. acknowledge additional 
support through SFB 631, SFB/TR 12, the Center for NanoScience (CeNS) Munich, and the German Excellence Initiative via the Nanosystems Initiative Munich (NIM).}

\end{document}